# NUMERICAL MODELING OF ELECTRODYNAMIC AGGREGATION OF MAGNETIZED NANODUST

V.S. Neverov, A.B. Kukushkin

NRC "Kurchatov Institute", Moscow 123182, Russia

**Abstract.** The recent results of applying the parallel numerical code SELFAS-3 to modelling of electrodynamic aggregation of magnetized nanodust are presented. The modelling describes evolution of a many-body system of basic blocks which are taken as strongly magnetized thin rods (i.e., one-dimensional static magnetic dipoles), with electric conductivity and static electric charge, screened with its own plasma sheath. The code provides continuous modelling of the following stages of evolution: (i) alignments of randomly situated solitary basic blocks in an external magnetic field and formation of stable filaments, (ii) percolation of electric conductivity in a random filamentary system, and electric short-circuiting in the presence of an external electric field, (iii) evolution of electric current profile in a filamentary network with a trend towards a fractal skeletal structuring.

## 1. Introduction

We continue analysing the capability of the model which has been suggested to explain the observed unexpected longevity of filaments, and their networks, in the high-current electric discharges (see [1] and references therein). This hypothesis predicted the macroscopic fractal structures with basic topological building block of tubular form (with presumably carbon nanotube (CNT) at the nanometer length scales), which is successively self-repeated at various length scales (see also the surveys [2,3] and web pages [4,5]).

The possibility of self-assembling of a fractal filamentary structure from a magnetized electroconductive nanodust was studied in [6-13]. We use the 3-D numerical model for a many body system of strongly magnetized thin rods (i.e. 1D static magnetic dipoles). Each block possesses the longitudinal electric conductivity and the electric charge, statically screened with its own plasma sheath. This approach has been implemented in the numerical code SELFAS-3 which is capable of parallel computing of electrodynamic aggregation in a many-body system of the above basic blocks. The code provided a continuous modelling of a transition between the following states: randomly situated ensemble of solitary basic blocks; electric current-carrying filamentary system; restructured filamentary network with a trend towards a fractal skeletal structuring. The latter trend was illustrated in [13] with generation of a bigger magnetic dipole in (i) initially homogeneous random ensemble between the biased electrodes in the presence of a plasma electric current filament and (ii) random ensemble around a straight linear nanodust filament with inhomogeneous distribution of the trapped magnetic flux along the filament [13].

Further substantial development of the hypothesis for a fractal macroscopic skeleton, which repeats the CNT structure at larger length scales, was suggested in [14] and continued with a series of papers by several groups, where mechanical and electrophysical properties of this hypothetical, virtually-assembled nanomaterial, named in [14] as "super carbon nanotube", have been studied theoretically with various numerical methods (see, e.g., most recent paper [15]).

Here we continue studying theoretically the ways to *fabricate* a wide class of fractal skeletal nanomaterial (not as ideal fractal as super CNT but still a fractal) via *electrodynamic*

*aggregation* of the above mentioned basic blocks with a strong contribution of internal self-organization processes (*self-assembling*).

## 2. Recent results on electrodynamic aggregation of magnetized nanodust

To illustrate the progress in the modeling of self-assembling processes we present an example of the evolution of a many body system of basis blocks in the case of a filamentary structure between two biased electrodes in the presence of external homogeneous magnetic field, directed along Z-axis, and of isotropic influx of solitary nanoparticles at a constant total rate. The results are shown in figures 1-12.

We use the same notations as in [13] where a short description of the code SELFAS-3 is given. Particular values of some parameters is as follows. Block's inverse aspect ratio $D/L = 0.06$, transition radius $r^* = D$ (tube's diameter), electric screening radius $R_D = 0.8L$, brake coefficients for tip-tip collision, $k_{br} = 100\ k_{br0}$, and for brake in an ambient medium, $M_{br} = 1.5\ k_{br0}$ (see Eq. (13) in [13]), $\delta_{J0} = 0.45$ (Eq. (1) in [13]), external Z-axial magnetic field $B_{ext} = 0.75\ B_0$ (Eq. (11) in [13]).

General view on the system is shown in figures 1 and 2. It is seen that formation of a skeletal structure around initially short-circuited few filaments takes plays due to a gradual enrichment of the skeleton and alignment of newly coming solitary blocks in the external and filaments-produced magnetic field. The 2D pictures for the evolution of particle density in the filamentary structure and current density are given in figures 3-5, while magnetic field is shown in figures 6 and 7. These figures present the distributions averaged over longitudinal (Z) direction of the entire filamentary structure. The 1D, radial profiles of the above parameters (particle and current density, and magnetic field) are shown in figures 8-10. Here, the above 2D distributions are averaged over azimuthal angle with respect to an axis of approximate symmetry. An important illustration of the system's dynamics is given in figures 11 and 12 where radial profiles of major electrodynamic forces are presented. It is seen that the quasi-stationary state in the system is determined by the force balance which essentially differs from that in the conventional dusty plasmas defined as a plasma with heavy dust particles with a strong electric charge. In particular, attraction of uncompensated magnetic dipoles within the electric current filaments to regions of a stronger magnetic field (Fig. 11(d)) may be a dominant component of the force balance counter-balanced by the strong Coulomb repulsion of blocks at close distances (Fig. 11(b)). In contrast, mutual interaction of electric currents through the filaments is compensated to an extent which roughly corresponds to a force-free equilibrium in an electric current-carrying plasmas (relative smallness of Ampère's force, Fig. 11(a)).

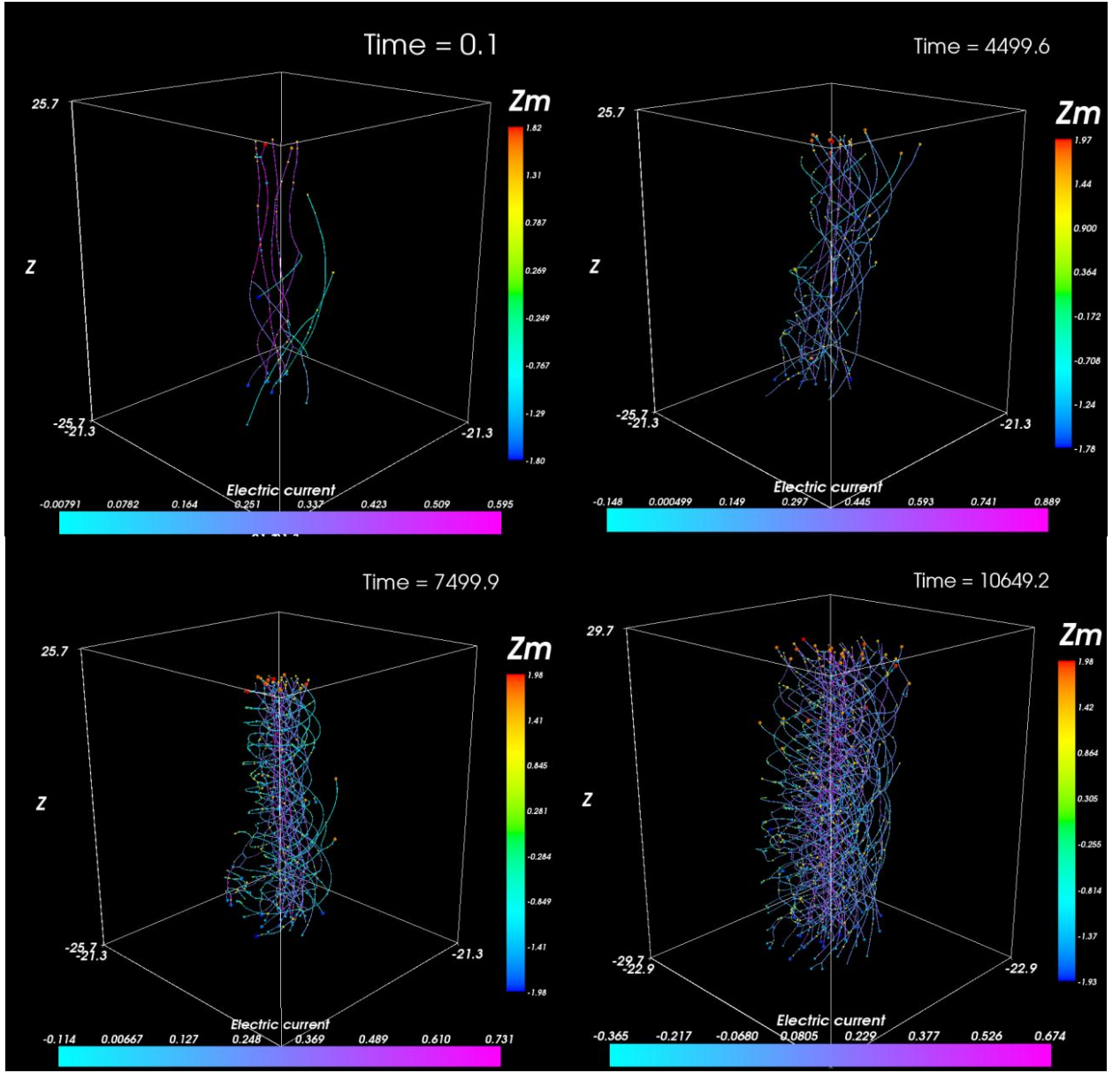

**Fig. 1.** Time evolution of a filamentary structure between two biased electrodes in the presence of external homogeneous magnetic field, directed along Z-axis, and of isotropic influx of solitary nanoparticles (at a total rate $\sim 1/t_0$). Time, electric current through the filaments and effective magnetic charge (i.e. magnetic dipole strength of individual block) are counted, respectively, in units of $t_0$, $J_0$, and $Z_{M0}$ defined by Eqs. (10) and (11) in [13]. Total number of particles trapped in the skeletal structure at maximal time is $\sim 4 \; 10^3$.

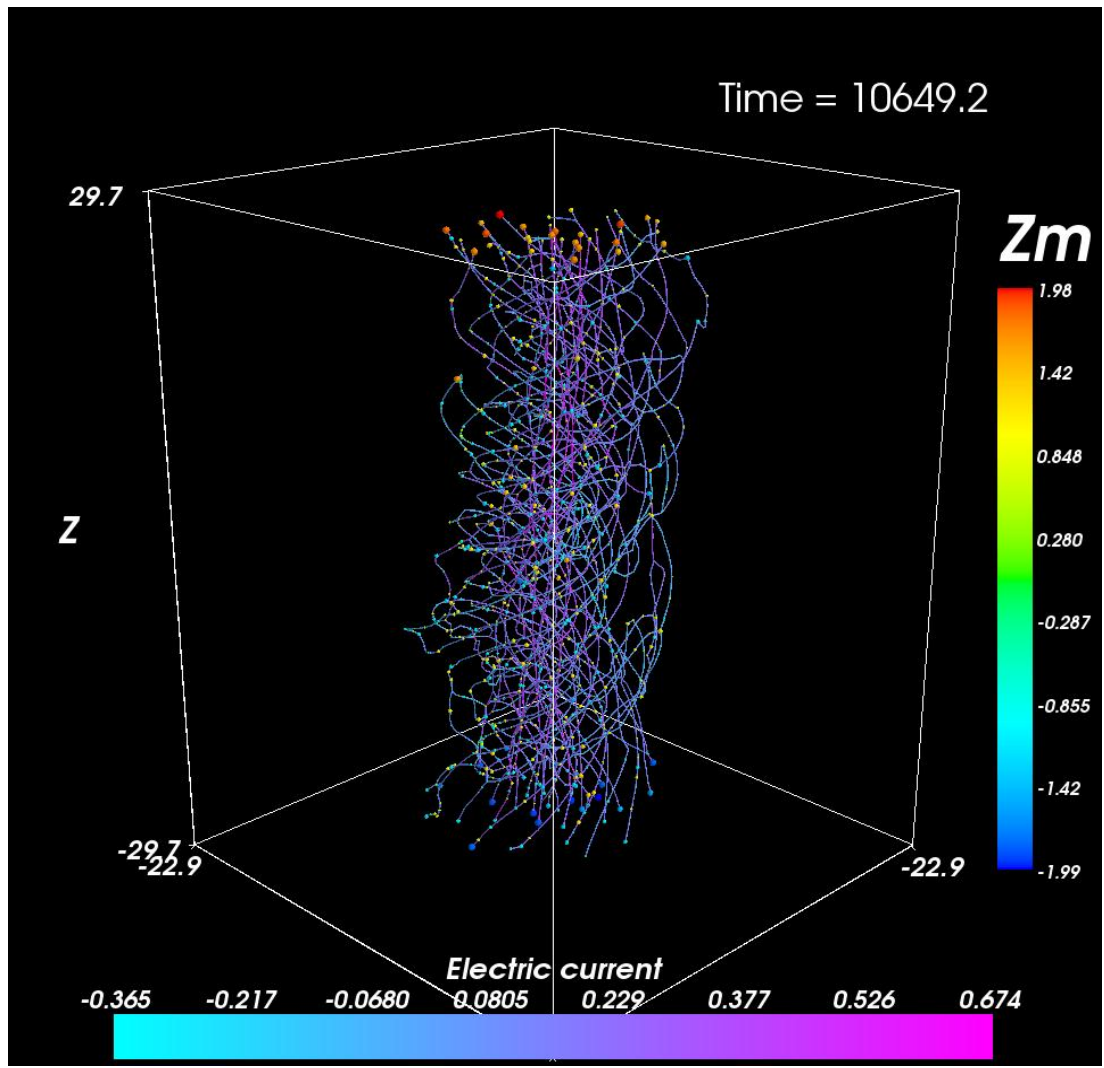

**Fig. 2.** The view of the filamentary structure in Fig. 1 at the end of modeling, shown without "dead ends", i.e. only the filaments with non-zero electric current are shown.

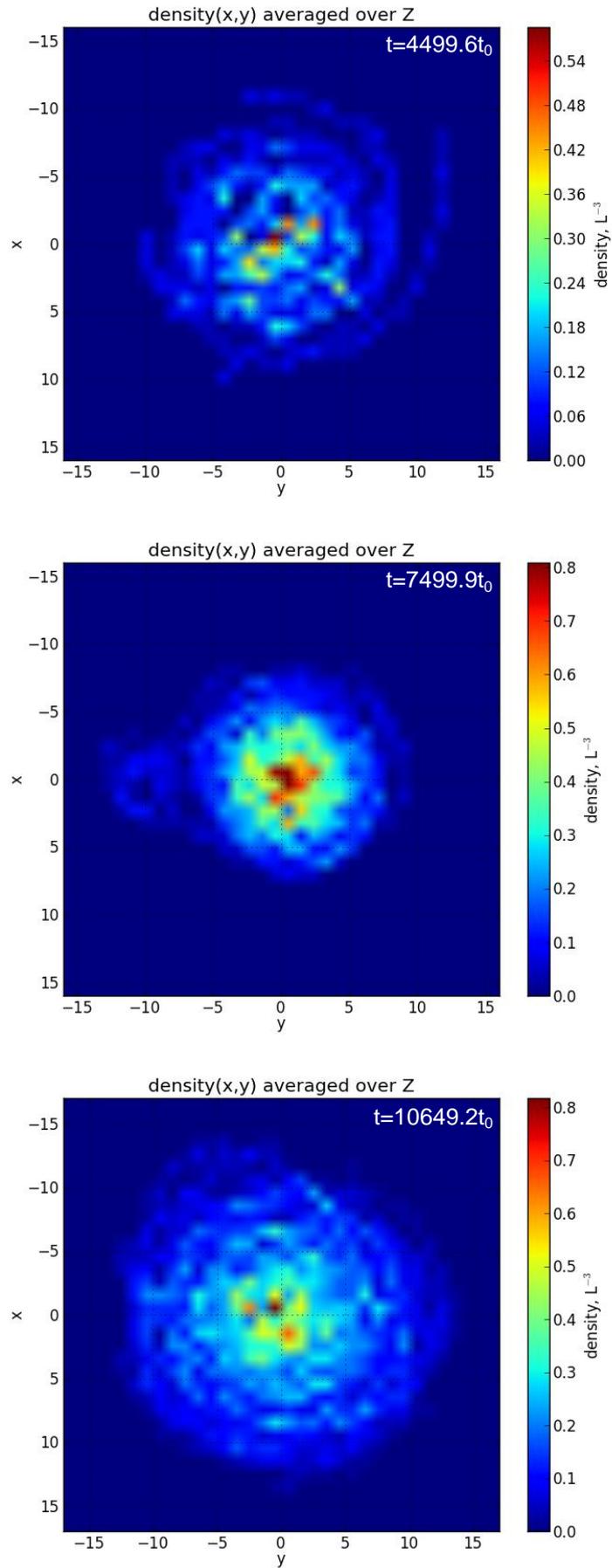

**Fig. 3.** Evolution of 3D density of basic blocks (in units $L^{-3}$, L is the length of 1D nanoparticle) in the filamentary structure, obtained by averaging over Z coordinate.

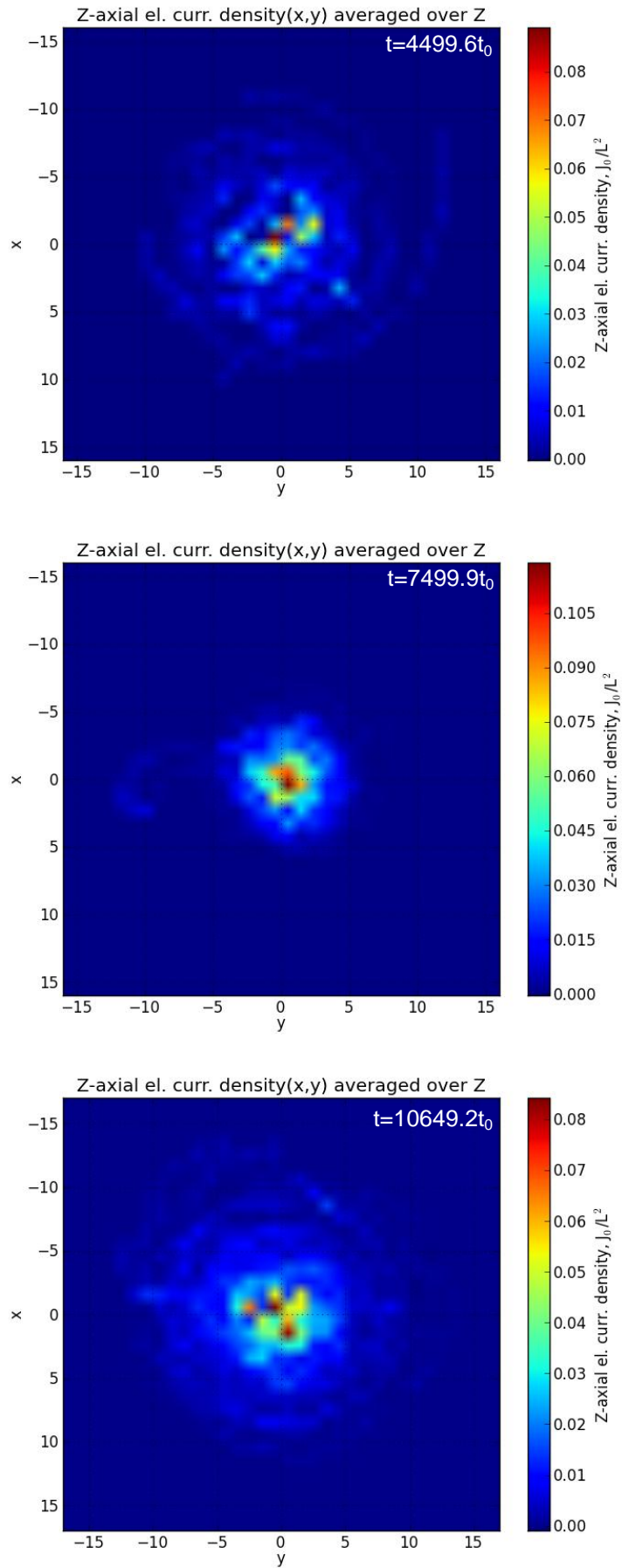

**Fig. 4.** Evolution of the density of longitudinal (i.e. Z-directed) electric current through the filaments (in units $J_0/L^2$), obtained by averaging over Z coordinate.

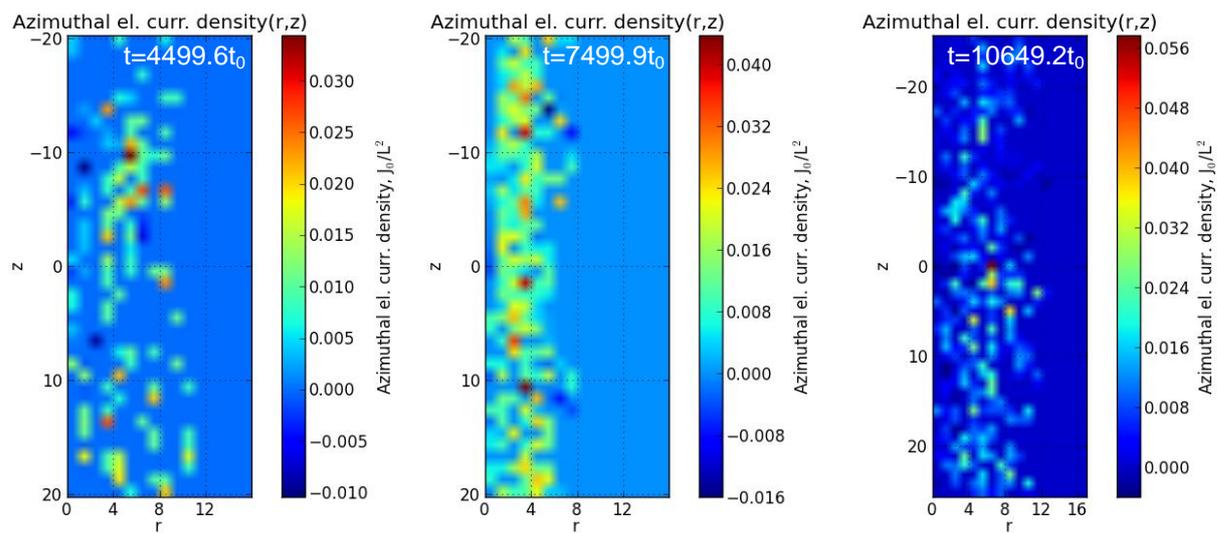

**Fig. 5.** Evolution of the density of azimuthal electric current density through the filaments (in units $J_0/L^2$), obtained by averaging over azimuthal angle.

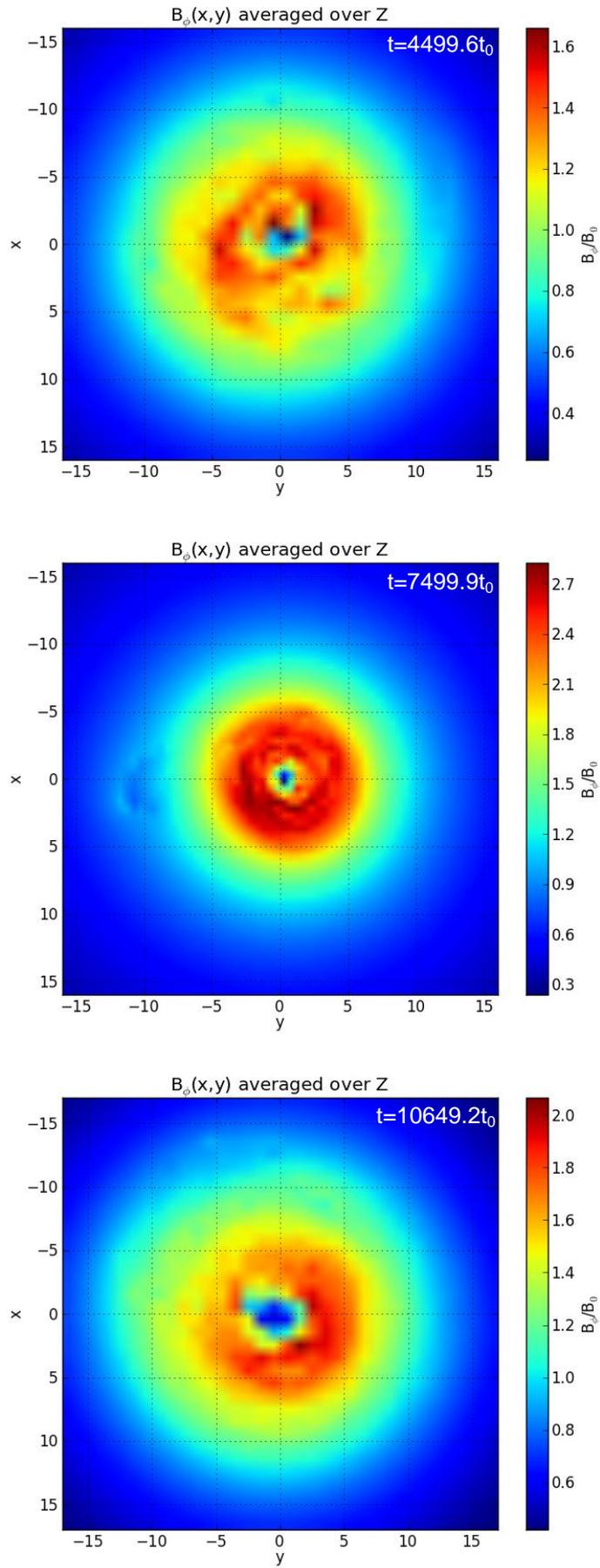

**Fig. 6.** Evolution of azimuthal component of magnetic field, in units $B_0$, produced by longitudinal electric current through the filaments.

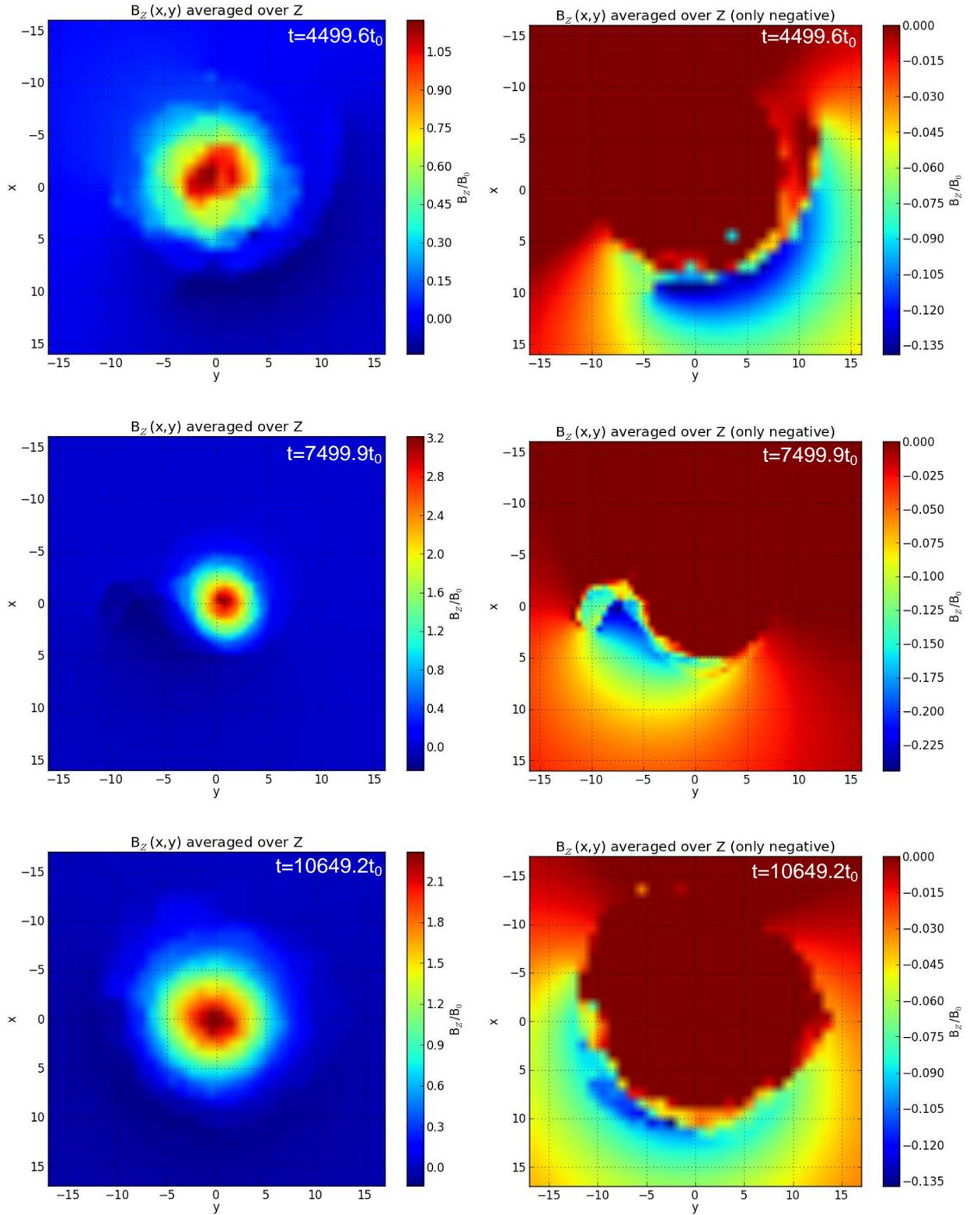

**Fig. 7.** Evolution of the longitudinal (Z) component of magnetic field, in units $B_0$, produced by azimuthal electric current through the filaments. In the right column regions with negative values are shown in more detail whereas all positive values are shown in brown.

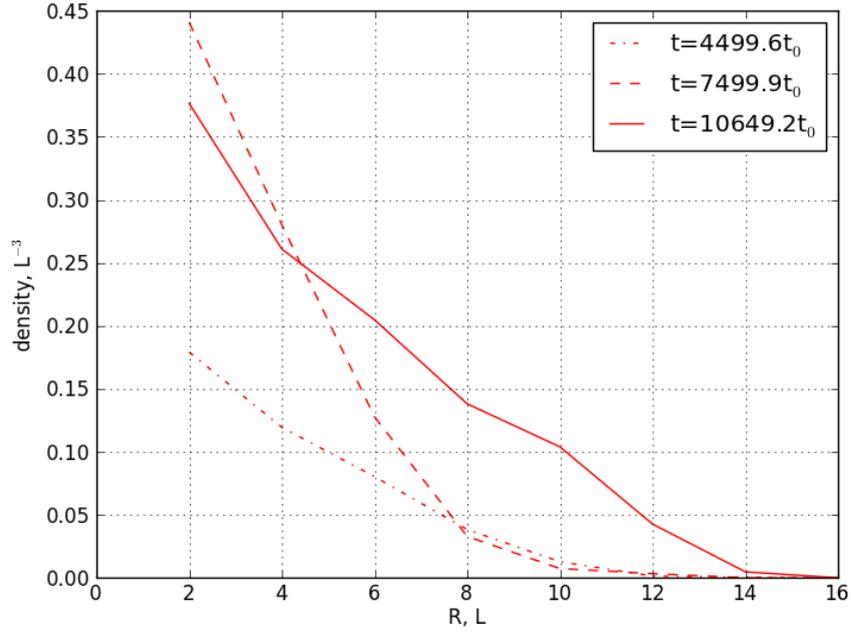

**Fig. 8.** Radial profiles of density of basic blocks (averaged over Z direction and azimuthal angle) at times $t = 4499.6\ t_0$ (dash-dot), $t = 7499.9\ t_0$ (dashed) and $t = 10649.2\ t_0$ (solid).

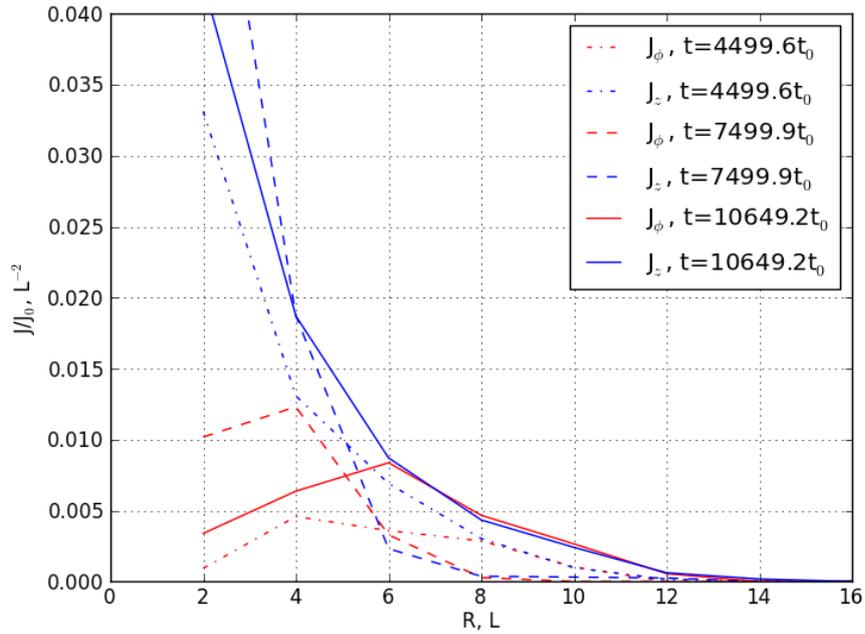

**Fig. 9.** Radial profiles of electric current density through the filamentary structure (averaged over Z direction and azimuthal angle) at times $t = 4499.6\ t_0$ (dash-dot), $t = 7499.9\ t_0$ (dashed) and $t = 10649.2\ t_0$ (solid). Red – azimuthal ($\phi$) component, blue – longitudinal (Z) component.

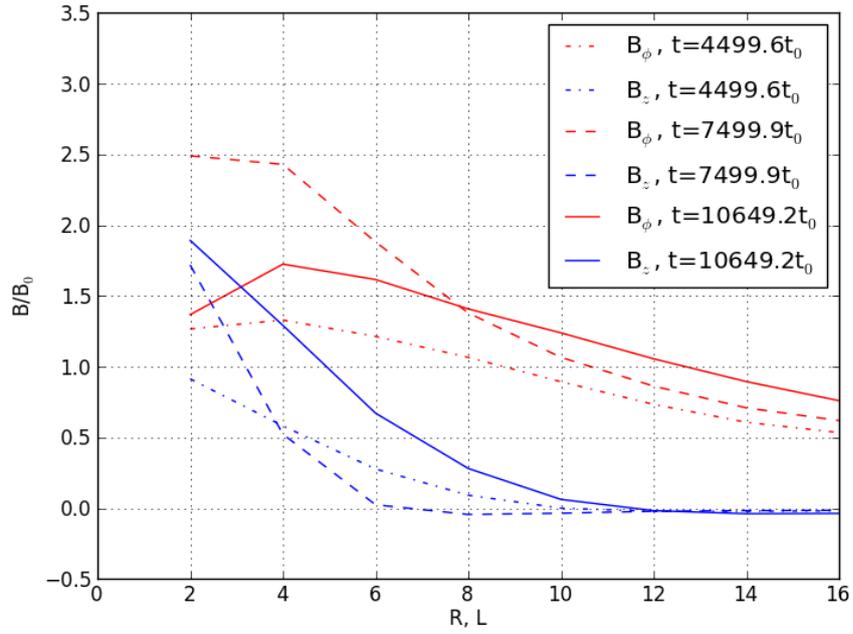

**Fig. 10.** Radial profiles of magnetic field components (averaged over Z direction and azimuthal angle) produced by electric currents through the filament structure at times $t = 4499.6\ t_0$ (dash-dot), $t = 7499.9\ t_0$ (dashed) and $t = 10649.2\ t_0$ (solid). Red − azimuthal ($\phi$) component, blue − longitudinal (Z) component.

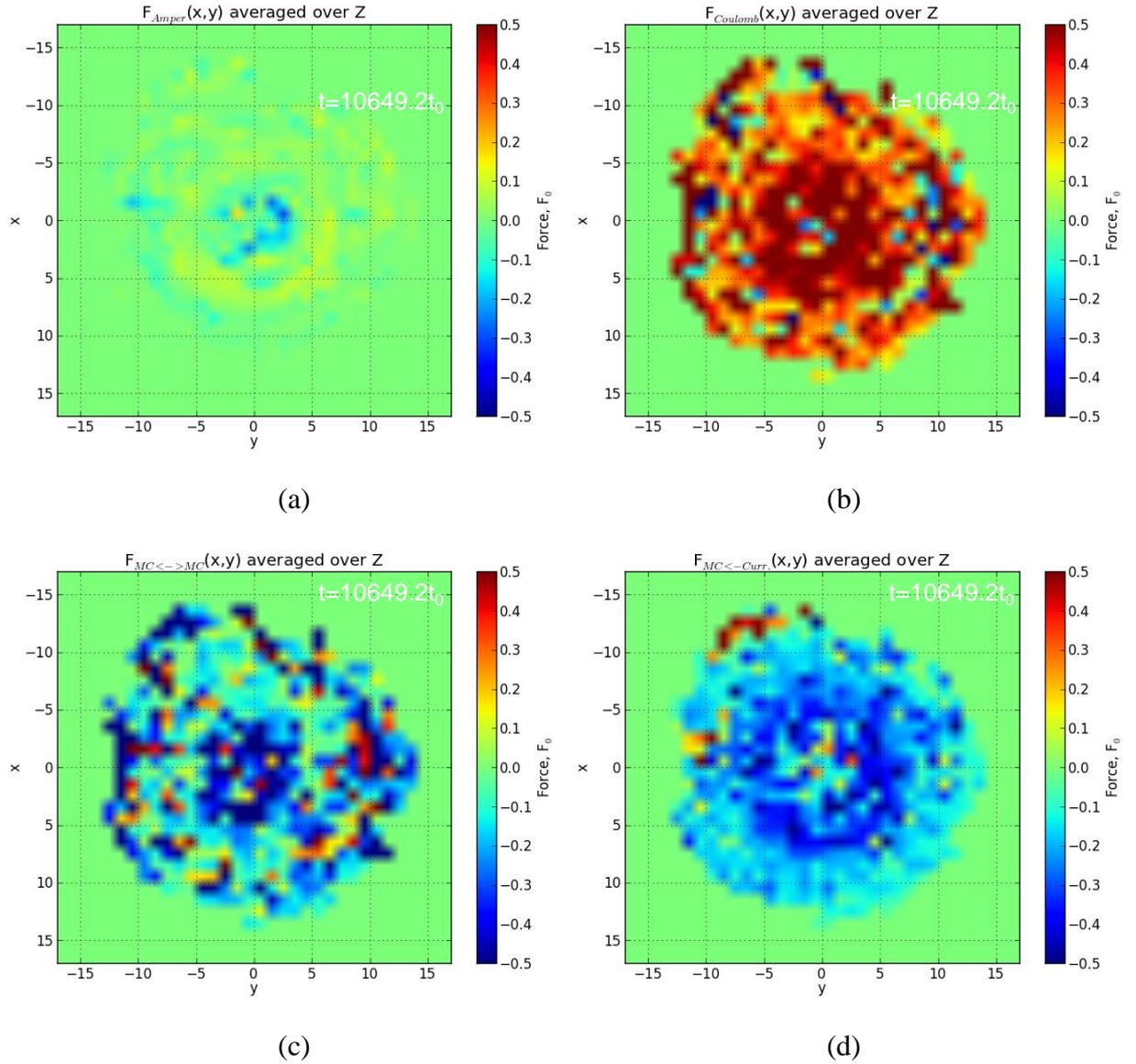

(a)

(b)

(c)

(d)

**Fig. 11.** Radial forces (averaged over Z coordinate) acting at a basic block at the time $t = 10649.2\ t_0$. "Ampere" (a) – interaction of electric currents through filaments composed of basic blocks with each other and with external magnetic field, "Coulomb" (b) – electric repulsion, "MC<–>MC" (c) – interaction of magnetic dipoles ("magnetic" elasticity of filaments), "MC<–Curr." (d) –action of electric current through the filaments on the magnetic dipoles.

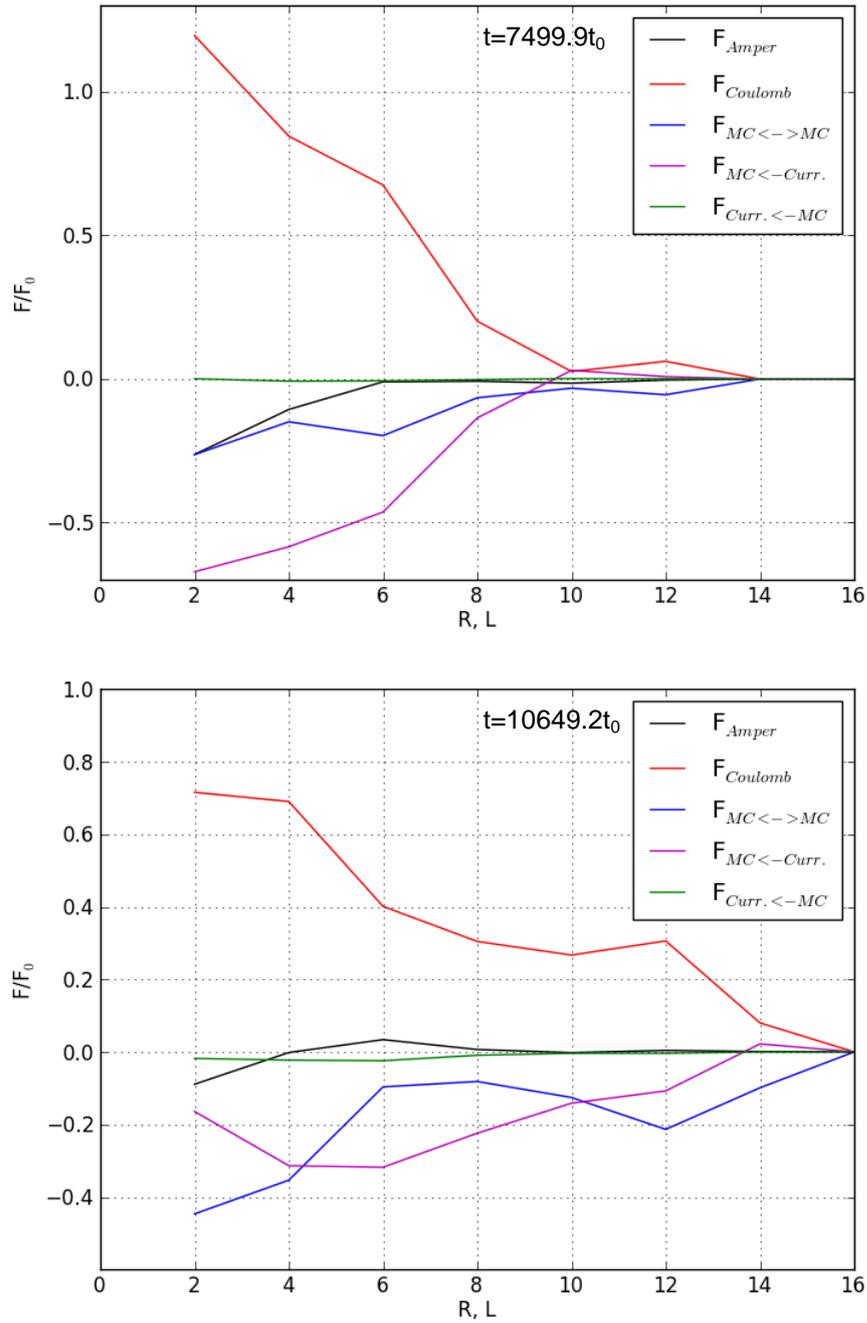

**Fig. 12.** Radial forces (averaged over Z direction and azimuthal angle) acting at a basic block at times $t = 7499.9\ t_0$ and $t = 10649.2\ t_0$. "Amper" – interaction of electric currents through filaments composed of basic blocks with each other and with external magnetic field, "Coulomb" – electric repulsion, "MC<–>MC" – interaction of magnetic dipoles (tension of filaments), "MC<–Curr." – force acting on the magnetic dipole from the electric current through the filaments composed of basic blocks, "Curr.<–MC" – force acting on the current through the basic block from magnetic dipoles.

## 3. Conclusions

1. The code SELFAS-3 [13] provides a continuous modelling of the following stages of evolution of the systems composed of strongly magnetized thin rods (i.e., one-dimensional static magnetic dipoles), with electric conductivity and static electric charge, screened with its own plasma sheath: (i) alignments of randomly situated solitary basic blocks in an external magnetic field and formation of stable filaments, (ii) percolation of electric conductivity in a random filamentary system, and electric short-circuiting in the presence of an external electric field, (iii) evolution of electric current profile in a filamentary network with a trend towards a fractal skeletal structuring.

2. The quasi-stationary states in such a system are determined by the force balance which essentially differs from that in the conventional dusty plasmas (heavy dust particles with a strong electric charge). In particular, attraction of uncompensated magnetic dipoles within the electric current filaments to regions of a stronger magnetic field may be a dominant component of the force balance.

3. Despite the full-scale numerical modeling of fractal skeletal structure formation is not feasible yet, the available results qualitatively suggest a way to fabricate a wide class of fractal skeletal nanomaterial via electrodynamic aggregation of magnetized nanodust: introduction of strong magnetic forces, inherent to a special type of nanodust, opens unprecedented opportunities for the percolation processes and skeletal structuring requested for various practical applications.

## Acknowledgments

The authors are grateful to I.B. Semenov, N.L. Marusov, V.A. Rantsev-Kartinov, P.V. Minashin (NRC "Kurchatov Institute"), A.P. Afanasiev, M.A. Posypkin, A.S. Tarasov, V.V. Voloshinov (Institute for System Analysis RAS), for helpful discussions.

This work is supported by the Russian Foundation for Basic Research (project RFBR 09-07-00469).